*Extended free-energy functionals for achiral and chiral ferroelectric nematic liquid crystals*

Yu Zou[1,2], Satoshi Aya[1,2]*

[1] *South China Advanced Institute for Soft Matter Science and Technology (AISMST), School of Emergent Soft Matter. South China University of Technology, China.*
[2] *Guangdong Provincial Key Laboratory of Functional and Intelligent Hybrid Materials and Devices, South China University of Technology, Guangzhou 510640, China.*

*Abstract*

Polar nematic liquid crystals are new classes of condensed-matter states where the inversion symmetry common to the traditional apolar nematics is broken. Establishing theoretical descriptions for the novel phase states is an urgent task. Here, we develop a Landau-type mean-field theory for both the achiral and chiral ferroelectric nematics. In the polar nematic states, the inversion symmetry breaking adds two new contributions: an additional odd elastic term (corresponding to the flexoelectricity in symmetry) to the standard Oseen-Frank free energy and an additional Landau term relating to the gradient of local polarisation. As a general necessity, the coupling between the scalar order parameter and polarisation order is further considered. In the chiral and polar nematic state, we reveal that the competition between the twist elasticity and polarity dictates effective compressive energy arising from the quasi-layer structure. The polarisation gradient is an essential term for describing the ferroelectric nature of the systems. The approaches provide theoretical foundations for testing and predicting polar structures in emerging polar liquid crystals.

*Introduction*

After the first discovery of ferroelectric liquid crystal (LC) states in the smectic C* phase several decades ago[1], LC research now extends to include mysterious polar nematic (N) LCs with high fluidity and strong polarisation[2-7]. In the traditional N state (Fig. 1a), the orientation of molecules in an ensemble is characterized by a head-to-tail equivalent unit vector called the director $\mathbf{n}$ ($\equiv -\mathbf{n}$). For example, like the simplest rod-shaped NLC molecule 4'-Pentyl-4-biphenylcarbonitrile (5CB), though the constituent molecular entity is polar along the long axis of molecules, an easy dimerization process prevents the system from forming polar LC phases at the macroscopic scale. The emerging polar LCs, such as the antiferroelectric splay nematic (SN; Fig. 1b)[8-10], achiral ferroelectric nematic (N$_F$; Fig. 1c)[3], chiral ferroelectric nematic (or helielectric nematic; HN*)[11, 12], heliconical ferroelectric nematic[13], and ferroelectric smectic-A (SmA$_F$)[14-16] states, belong to a new category of LC phases that do allow some local residual polarisations. Especially in the N$_F$, long-pitch HN*, and SmA$_F$ states, the macroscopic inversion symmetry breaking occurs, i.e., $\mathbf{n} \neq -\mathbf{n}$[17], resulting in spontaneous polarisation. Several studies have shown potential microscopic origins of spontaneous polarity and revealed the roles of steric and electric-field effects on the stability of polar states[18-23]. In terms of symmetry, the N state and the polar counterpart, the N$_F$ state, belong to the $D_{\infty h}$ and $C_{\infty v}$ point groups, respectively[7]. Practically, the realization of the polar states is made by introducing strong dipole moments to rod-shaped molecules[4, 24]. Compared with traditional bent-core LCs[25], rather than having their dipole moments perpendicular to the long axis of bent-shaped molecules, most of the current synthetic N$_F$ materials carry strong dipoles along the long axis of rod-shaped molecules. The head-to-tail inequivalent symmetry gives rise to many unprecedented properties, such as large values of the second-order nonlinear optical coefficient[4, 11], dielectric anisotropy[3], and spontaneous polarisation (up to 6 µC/cm$^2$)[3, 26].

When an LC system becomes head-to-tail asymmetric, the description of the elastic properties would differ from the traditional apolar LCs due to the symmetry of $\mathbf{n} \neq -\mathbf{n}$. Intuitively, the symmetry requires the typical Ossen-Frank elastic free-energy modified with additional odd-order terms. Meanwhile, according to the classical argument by Meyer[1], a polar order along the director $\mathbf{n}$ (i.e., polarisation vector $\mathbf{P} = P_0 \mathbf{n}$) can couple to the splay deformation, called flexoelectricity. $P_0$ is the magnitude of the polarisation. The strong molecular polarity used for the emerging polar states ensures much stronger flexoelectric coupling than the traditional LCs. Meanwhile, the depolarisation field effect arising from the divergence of the polarisation field tends to suppress the splay deformation of $\mathbf{P}$. While the splay deformation would be strongly prohibited by the depolarisation field effect in the matured N$_F$ state[27], the flexoelectricity would dominate at the onset of the N$_F$ state[28] or during its development from apolar phases[8-10]. The competition between the two effects, as well as the nematic elasticity, would give rise to a rich variety of new LC phases and topologies. Indeed, though very short from the discovery of the N$_F$ phase, new polar phases like HN* and SN[8, 12] and unique polarisation topologies[7, 28-31] were successively found thanks to these competitions.

As extensions of the traditional NLC theory[32], numerous mean-field models for polar LCs have been considered so far. For example, Alenka et al.[8] incorporated the flexoelectric coupling of polarisation **P** and a gradient term of *Q*-tensor ($\nabla \mathbf{Q}$) into the Landau–de Gennes free energy functional, successfully explaining a flexoelectricity-driven elastic softening of splay deformation near the transition from the N to the splay nematic (SN) state. Rosseto et al.[10] re-examined the Landau functional proposed by Alenka et al.[8], and theoretically compared the energy stability of one- (single-splay) and two-dimensional modulations of the polarisation field under the condition of strong flexoelectric coupling. Kats et al.[33] analysed the stability of a uniform $N_F$ state in both cases of the polarisation **P** parallel (easy-axis case) or perpendicular (easy-plane case) to the director **n**. They specified that the splay nematic phases reported in Ref. [8, 10] are more likely to form in the easy-axis case compared to the easy-plane case. The uniform $N_F$ state loses its stability mainly attributed to the flexoelectric coupling. Caimi et al.[5] addressed the determinative role of the depolarisation effect on realigning the polarisation field in the $N_F$ phase. Our previous studies[7, 29] clarified the dominant roles of polar interactions, including polarisation gradients, flexoelectric effects, and depolarisation effects, coupled with specific spatial confinement conditions, in determining the polarisation field in the $N_F$ and HN* states. From the microscopic perspective, Madhusudana[19] modelled highly polar LC molecules as cylindrical rods with longitudinal charge density waves, demonstrating that electrostatic interactions favour parallel over antiparallel structures under particular conditions. Therefore, the $N_F$ state can be stabilized. In the previous studies, how the spontaneous breaking of the inversion symmetry leads to a natural modification of the nematic elastic properties is not clear. In addition, how the extended free-energy functionals would benefit to or connect to experimental measurement is not given.

In this article, we derive the elastic and Landau free-energy functionals from the perspective of symmetry for the $N_F$ and HN* states and establish the free-energy descriptions for the systems. We compare the properties between the N and $N_F$ states and between the apolar chiral nematic (N*) and HN* states. The theoretical foundation makes it possible to explore potential polar states and topologies more accurately and could be extended for polar smectic states.

*Result and Discussion*

*Free-energy functional for $N_F$ state* – Let us first discuss how the Oseen-Frank elastic free-energy functionals in the typical apolar N state would be revised in the $N_F$ state. We consider the director **n** at the origin aligns along the *z*-axis. We assume that the director field exhibits a continuous variation, and **n** deviates from the $z$-axis at a position slightly away from the origin defined by a displacement vector $\delta \mathbf{r} = (\delta x, \delta y, \delta z)$, The director is expressed as

$$\mathbf{n}(\delta \mathbf{r}) = \begin{pmatrix} n_x \\ n_y \\ n_z \end{pmatrix} = \begin{pmatrix} \frac{\delta n_x}{\delta x}\delta x + \frac{\delta n_x}{\delta y}\delta y + \frac{\delta n_x}{\delta z}\delta z \\ \frac{\delta n_y}{\delta x}\delta x + \frac{\delta n_y}{\delta y}\delta y + \frac{\delta n_y}{\delta z}\delta z \\ 1 + O(|\delta \mathbf{r}|^2) \end{pmatrix}$$
$$= \begin{pmatrix} a_1 \delta x + a_2 \delta y + a_3 \delta z \\ a_4 \delta x + a_5 \delta y + a_6 \delta z \\ 1 + O(|\delta \mathbf{r}|^2) \end{pmatrix}. \tag{1}$$

$a_1$ and $a_5$ describe the splay deformation in the $xz$ and $yz$ planes, respectively. $a_2$ and $-a_4$ the twist deformation, and $a_3$ and $a_6$ the bend deformation (Figs. 1d,i). At least more than one of $a_i$ ($i = 1, 2, 3, 4, 5, 6$) should be nonzero to make an orientational deformation. Learn from Hooke's Law, that the free-energy density arising from the orientational deformation, $f$, is represented by a power series of the gradient of the director field, $a_i$. Up to the second power,

$$f_{\text{elastic}} = K_i a_i + \frac{K_{ij} a_i a_j}{2} \ (i, j = 1, 2, 3, 4, 5, 6; \ K_{ij} = K_{ji}), \tag{2}$$

by Einstein's summation convention. $K_i$ and $K_{ij}$ are the nematic elastic constants for the first-order and second-order power series expansions. When particular coefficients remain under a specific symmetry, the corresponding elastic deformation modes would be allowed.

To derive the free-energy density expression for the $N_F$ state, it is convenient to argue the energy invariance under coordinate transformation. Let us consider a coordinate transformation: $(x, y, z) \to (x', y', z')$. The orientational deformation $a_i$ ($i = 1, 2, 3, 4, 5, 6$) in the new coordination re-reads $a_i'$ ($i = 1, 2, 3, 4, 5, 6$). Then, the free-energy density can be also re-written as

$$f'_{\text{elastic}} = K_i a_i' + \frac{K_{ij} a_i' a_j'}{2}. \tag{3}$$

Recall that, in the apolar N state, the $D_{\infty h}$ symmetry permits the director field to cover half of the $S^1/Z_2$ order parameter space (Figs. 2a-d). The rest half is identical to the first half. In the $N_F$ state, the $C_{\infty v}$ symmetry allows the polarisation field to wrap a full $S^2$ order parameter space (Figs. 2a-d). These directly mean that the symmetry operations in $D_{\infty h}$ and $C_{\infty v}$ point groups guarantee the invariance upon $(x, y, z) \to (x', y', z')$ for the N and $N_F$ states, respectively. Meanwhile, Eqs. (2) and (3) should give the same free-energy value upon arbitrary coordinate transformations in the corresponding point groups. In the following, we show four cases of coordinate transformations to find nonzero $K_i$ and $K_{ij}$ coefficients in the $N_F$ state: (1) $\pi$ rotation about $z$-axis (Fig. 2a); (2) $\pi/2$ rotation about $z$-axis (Fig. 2b); (3) $\pi/4$ rotation about $z$-axis (Fig. 2c); (4) $\pi$ rotation about $x$-axis (Fig. 2d).

In case (1), $(x, y, z) \to (x', y', z') = (-x, -y, z)$ leads to the corresponding director variation $\mathbf{n} = (n_x, n_y, n_z) \to \mathbf{n}' = (n'_x, n'_y, n'_z) = (-n_x, -n_y, n_z)$. According to the definitions of $a_i$ (Eq. (1)), we get $a_1' = a_1$, $a_2' = a_2$, $a_3' = -a_3$, $a_4' = a_4$, $a_5' = a_5$, $a_6' = -a_6$. To ensure Eqs. (2) and (3) identical, $K_i$ and $K_{ij}$ coefficients for terms under $a_i' = -a_i$ and $a_i' a_j' = -a_i a_j$ conditions should vanish. Therefore, we obtain $K_3 = K_6 = K_{13} = K_{16} = K_{23} = K_{26} = K_{34} = K_{35} = K_{46} = K_{56} = 0$. Following the same logic, we get $K_4 = -K_2$, $K_1 = K_5$, $K_{22} = K_{44}$, $K_{11} = K_{55}$, $K_{33} = K_{66}$, $K_{25} = -K_{14}$, $K_{45} = -K_{12}$, $K_{36} = 0$ for case (2) and $K_{15} = K_{11} - K_{22} - K_{24}$, $K_{14} = -K_{12}$ for case (3). These symmetry operations for cases (1)-(3) result in the same descriptions for the N and $N_F$ states, as easily understood by the same director variations in the order parameter space (Figs. 2a-c). In case (4), the difference between the N and $N_F$ states appears. In the N state, $(x, y, z) \to (x', y', z') = (x, -y, -z)$ leads to $\mathbf{n} = (n_x, n_y, n_z) \to \mathbf{n}' = (n'_x, n'_y, n'_z) = (n_x, -n_y, -n_z) \equiv (-n_x, n_y, n_z)$ (Fig. 2d), resulting in $K_1 = 0$, $K_{12} = 0$. On the other hand, in the $N_F$ state, $(x, y, z) \to (x', y', z') = (x, -y, -z)$ leads to $\mathbf{n} = (n_x, n_y, n_z) \to \mathbf{n}' = (n'_x, n'_y, n'_z) = (n_x, -n_y, -n_z)$ (Fig. 2d), then we arrive $K_2 = K_4 = 0$, $K_{12} = 0$. These four situations enable us to get a complete set of expressions about $K_i$ and $K_{ij}$ coefficients in the form of the matrix for the $N_F$ state:

$$\boldsymbol{K}_i = \begin{pmatrix} K_1 \\ K_2 \\ K_3 \\ K_4 \\ K_5 \\ K_6 \end{pmatrix} = \begin{pmatrix} K_1 \\ 0 \\ 0 \\ 0 \\ K_1 \\ 0 \end{pmatrix} \tag{4}$$

$$\boldsymbol{K}_{ij} = \begin{pmatrix} K_{11} & 0 & 0 & 0 & K_{15} & 0 \\ 0 & K_{22} & 0 & K_{24} & 0 & 0 \\ 0 & 0 & K_{33} & 0 & 0 & 0 \\ 0 & K_{42} & 0 & K_{22} & 0 & 0 \\ K_{51} & 0 & 0 & 0 & K_{11} & 0 \\ 0 & 0 & 0 & 0 & 0 & K_{33} \end{pmatrix} \tag{5}$$

Worth noting that while the $K_{ij}$ matrix is common to the N state, the $K_i$ matrix differs from that in the N state, where no odd term exists ($K_2$ vanishes because of the invariance upon $(x, y, z) \rightarrow (x', y', z') = (-x, -y, -z)$). The nonzero $K_1$ and $K_4 = K_1$ means that the odd terms of $\delta n_x / \delta x$ and $\delta n_y / \delta y$ regarding 'directional' splay deformation modes survive. This makes a significant distinction to the N state, where the lowest order for the splay deformation is $(\delta n_i / \delta i)^2$, saying that any splay deformation is free-energy costing. In the $N_F$ state, the splay deformation could be free-energy favourable in a spontaneous manner when the odd term (say "positive" splay) becomes negative. This scenario, naturally derived from the symmetry argument, is consistent with proceeding theoretical predictions[28], corresponding to the flexoelectric effect. Now, we have the elastic free-energy description for the $N_F$ state as:

$$f_{\text{elastic}} = K_1(a_1 + a_5) + \frac{K_{11}(a_1 + a_5)^2}{2} + \frac{K_{22}(a_2 - a_4)^2}{2} + \frac{K_{33}(a_3^2 + a_6^2)}{2} - (K_{22} + K_{24})(a_1 a_5 - a_2 a_4) \quad (6)$$

The last term does not contribute to the bulk free energy, corresponding to the saddle-splay deformation. The bulk elastic free energy function can also be turned into the expression:

$$f_{\text{elastic}} = K_1(\nabla \cdot \mathbf{n}) + \frac{K_{11}}{2}(\nabla \cdot \mathbf{n})^2 + \frac{K_{22}}{2}[\mathbf{n} \cdot (\nabla \times \mathbf{n})]^2 + \frac{K_{33}}{2}[\mathbf{n} \times (\nabla \times \mathbf{n})]^2 \quad (7)$$

The leading term that couples to the divergence of the director field indicate a preference for a "positive" splay (so $K_1 < 0$), where the free-energy density becomes negative. In real symmetry-broken systems, the polar symmetry can be given by introducing large dipole or asymmetric shape anisotropy (e.g., pearl-shape)[34]. As a result, the first-order odd term can be considered as the flexoelectric interaction in the $N_F$ state as $-\gamma \mathbf{n}(\nabla \cdot \mathbf{n}) \cdot \mathbf{P}$ instead. $\gamma$ is a bare flexoelectric coefficient. $\mathbf{P}$ is the spontaneous polarisation with its magnitude $P_0$ and its direction parallel with director $\mathbf{n}$, i.e., $\mathbf{P} = P_0 \mathbf{n}$.

As well as the elastic energy, the stability of a polar state should be determined by the Landau free energy. An important feature of the $N_F$ state is that the ferroelectric ordering dislikes any distortion of the polarisation field, which is usually described in the Ising model by setting a negative coefficient for the Hamiltonian $H = J\mathbf{P}_i \mathbf{P}_j$; ($\mathbf{P}_i$ and $\mathbf{P}_j$ are the adjacent polarisation and $J < 0$ for ferroelectrics)[35]. This feature means that we must include the energy cost for the inhomogeneous polarisation field. This operation can be made naturally by expanding the free-energy density in powers of the polarisation and its gradient. We choose the reference point of the free-energy density for the unpolarized and homogeneous polarisation field as zero in free energy. Then, to the lowest orders, we can write the Landau free-energy density as:

$$f_{\text{Landau}} = \frac{A|\mathbf{P}|^2}{2} + \frac{B|\mathbf{P}|^4}{2} + \frac{h}{2}|\nabla \mathbf{P}|^2, \tag{8}$$

$$\nabla \mathbf{P} = \begin{pmatrix} \frac{\partial P_x}{\partial x} & \frac{\partial P_x}{\partial y} & \frac{\partial P_x}{\partial z} \\ \frac{\partial P_y}{\partial x} & \frac{\partial P_y}{\partial y} & \frac{\partial P_y}{\partial z} \\ \frac{\partial P_z}{\partial x} & \frac{\partial P_z}{\partial y} & \frac{\partial P_z}{\partial z} \end{pmatrix}. \tag{9}$$

$\nabla \mathbf{P}$ term includes the splay, twist and bend polarisation deformations. The first two terms of Eq. (8) are common to the descriptions for many ferroelectrics[35], including the traditional polar LCs like ferroelectric smectic-C* state[36, 37]. The third, i.e., the gradient term, corresponds to modifying the Landau-Devonshire expression and tends to prohibit polarisation deformations[35]. We note that the $|\nabla \mathbf{P}|^2$ term is necessary to judge if the system prefers the ferroelectric state. This is because the first two Landau terms only care about whether the system would stabilize a polar state that could be either macroscopically ferroelectric or antiferroelectric or even other deformed polar states.

Finally, we must also include the free-energy density arising from polar interactions. As discussed in the elastic terms, the additional first-order term of the director gradient physically corresponds to the flexoelectricity ($-\gamma \mathbf{n}(\nabla \cdot \mathbf{n}) \cdot \mathbf{P}$), which is essential to determine the polarisation alignment and topology in the $N_F$ and HN* states near the apolar-polar transition point[7, 8, 28]. The depolarisation charge effect is another crucial feature that should be introduced[5, 6, 27-30]. It tends to exclude the appearance of depolarisation charges that arise from splay deformations of the polarisation field. The effect is written as:

$$f_{\text{dep}} = -\frac{1}{2} \mathbf{P} \cdot \mathbf{E}_d. \tag{10}$$

$\mathbf{E}_d$ is the depolarisation field, which can be solved by using the Poisson's equation,

$$\nabla^2 \Phi = -\frac{\sigma_d}{\varepsilon}, \tag{11}$$

$$\mathbf{E}_d = -\nabla \Phi. \tag{12}$$

$\sigma_\mathrm{d} = -\nabla \cdot \mathbf{P}$ is the depolarisation charge density. $\varepsilon$ is the dielectric constant of the material. $\Phi$ is the depolarisation potential. Though, usually, $f_\mathrm{dep}$ cannot be solved analytically, the aforementioned polarisation works as a rough for accounting for the penalization of the splay deformation. Therefore, for analytical usage, one can choose to include the electrostatic interaction in the polarisation gradient energy by using with an additional electrostatic coefficient. While it is a rude approximation, it is satisfactory to some extent for reproducing some topological nature in the $N_F$ state in the previous report[33]. The essence of the depolarisation effect is the Coulomb interaction, which is a long-range interaction. On the other hand, the polarisation gradient can also be regarded as a semi-long-range interaction with a length order similar to the characteristic length of nematic because it tends to prevent discontinuous variation of the director field. Adding all the discussed terms up, we reach the free-energy functional for the $N_F$ state as

$$f_\mathrm{NF} = \frac{K_{11}}{2}[\nabla \cdot \mathbf{n}]^2 + \frac{K_{22}}{2}[\mathbf{n} \cdot (\nabla \times \mathbf{n})]^2 + \frac{K_{33}}{2}[\mathbf{n} \times (\nabla \times \mathbf{n})]^2$$
$$-\gamma \mathbf{n}(\nabla \cdot \mathbf{n}) \cdot \mathbf{P} + \frac{A|\mathbf{P}|^2}{2} + \frac{B|\mathbf{P}|^4}{2} + \frac{h}{2}|\nabla \mathbf{P}|^2 - \frac{1}{2}\mathbf{P} \cdot \mathbf{E}_\mathrm{d}. \quad (13)$$

However, this form can only analyse energy landscape under the constant scalar order parameter ($s$) and polarisation ($P_0$) condition, which would be problematic for conducting theoretical analyses and numerical simulations that allow variations of the scalar order parameter and the magnitude of the polarisation. The main issue here is the need for the coupling between the nematic director and polarisation field, i.e., $s$ and $P_0$ vary independently. In the simplest case where the dipolar interaction is the dominant factor for giving rise to the (polar) nematic order, it is expected that the scalar order parameter of the nematicity is proportional to the magnitude of the polarisation. It means that the polarisation magnitude should change accordingly once the nematic order parameter changes. Under this condition, a polar nematic phase appears directly from the isotropic phase without going into an apolar nematic phase. To account for this situation, we revise both the expressions of $\mathbf{n}$ and $\mathbf{P}$. The modification of $\mathbf{n}$ is similar to the Ericksen expressions [38-40], coupled to the scalar order parameter $s$: $\mathbf{N} \equiv s\mathbf{n}$. Since the variation of $s$ would linearly change the magnitude of the polarisation, we revise $\mathbf{P}$ as $\mathbf{P} \equiv P_0\mathbf{N} = sP_0\mathbf{n}$ (or more generally with a nonlinearity, $P_0\mathbf{N} = s^\gamma P_0\mathbf{n}$). $\gamma$ is the exponent. $P_0$ is a constant, expressing the saturation value of the spontaneous polarisation. When the shape anisotropy is also important and would compete with the dipolar anisotropy, some nonlinearity would arise. Unless the nonlinearity is significant, the linear approximation would be satisfactory. With these modifications, we reach the final form of the free-energy functional for the $N_F$ state as

$$f_{NF} = \frac{1}{2}s^2 K_{11}[\nabla \cdot \mathbf{n}]^2 + \frac{1}{2}s^4 K_{22}[\mathbf{n} \cdot (\nabla \times \mathbf{n})]^2$$
$$+ \frac{1}{2}s^4 K_{33}[\mathbf{n} \times (\nabla \times \mathbf{n})]^2 - \gamma s^3 P_0 \mathbf{n}(\nabla \cdot \mathbf{n}) \cdot \mathbf{n}$$
$$+ \frac{As^2 P_0^2 |\mathbf{n}|^2}{2} + \frac{Bs^4 P_0^4 |\mathbf{n}|^4}{2} + \frac{h}{2}P_0^2 |\nabla(s\mathbf{n})|^2$$
$$- \frac{1}{2}sP_0 \mathbf{n} \cdot \mathbf{E}_d \tag{14}$$

*Free-energy functional for HN* state* – Next, we move to discuss the free energy for the chiral ferroelectric state, i.e., the HN* state. As we already obtained the free-energy formulation for the $N_F$ state, the extended Oseen-Frank theory can be used to describe the HN* state by simply adding chiral terms. The equilibrium structures of the N* or HN* state are periodic twists around a helical axis (Figs. 3a and b). Here we assume that the helical axis orients along the *z*-axis. The periodicity of the twists is $2\pi/q_0$. Therefore, the equilibrium director field is written as $\mathbf{n}_0 = (\cos q_0 z, \sin q_0 z, 0)$. In the HN* state, a twist of a particular twisting sense is favoured by the system. We arrive at the elastic free energy:

$$f^*_{elastic} = \frac{s^2 K_{11}}{2}(\nabla \cdot \mathbf{n})^2 + \frac{s^4 K_{22}}{2}[\mathbf{n} \cdot (\nabla \times \mathbf{n}) + q_0]^2$$
$$+ \frac{s^4 K_{33}}{2}[\mathbf{n} \times (\nabla \times \mathbf{n})]^2 + sK_1(\nabla \cdot \mathbf{n}), \tag{15}$$

so that the HN* state with a pure twisted deformation of period $2\pi/q_0$ is at the equilibrium. For helical LCs (e.g., N* and HN*), the twists can also be treated as quasi-layers (Fig. 3c). Therein, an alternative mean-field free-energy theory, called the Lubensky-de Gennes form of the free-energy density[41], can be considered. The free energy reads

$$f_{L-G} = \frac{B}{2}\left[\frac{\partial u}{\partial z} - \frac{1}{2}\left(\frac{\partial u}{\partial x}\right)^2 - \frac{1}{2}\left(\frac{\partial u}{\partial y}\right)^2\right]^2 + \frac{K}{2}\left(\frac{\partial^2 u}{\partial x^2} + \frac{\partial^2 u}{\partial y^2}\right)^2 \tag{16}$$

The first term represents the compression energy originating from the quasi-layer nature of the elastic helix, penalizing the layer compression or dilation (or director bend). $B$ is the compression modulus for the quasi-layer and $u$ is the displacement of the quasi-layer. The second term describes the energy payment upon a bending of quasi-layers bending of the quasi-layers. $K$ is the layer bending (or effective director splay) modulus. This expression was already used in our previous report[29], where the polarity is decoupled from both the elastic and compressive terms.

To obtain a simpler polarity-elasticity coupled equation, we consider an effective elastic term and a compression term to comprise the free-energy density as:

$$f_{\text{L-G}} = \frac{\widetilde{K}}{2}(\nabla \cdot \chi)^2 + \frac{\widetilde{B}}{2}\left(\frac{p}{p_0} - 1\right)^2 \quad (17)$$

$\widetilde{K}$ and $\widetilde{B}$ are the effective elastic and compression coefficients, respectively. $\chi$ is a unit vector along the helical axis. $p$ and $p_0$ represent the actual and equilibrium pitch lengths for the N* or HN* systems, respectively. This expression is practical because real measurements probe the effective elasticity and compression modulus, where the contributions from the nematic elasticity and polar interactions are already included. Therefore, the central issue now is to derive the expression of $\widetilde{K}$ and $\widetilde{B}$, which should link to both the polarity and the nematic elasticity for polar states. To make a full comparison with the traditional apolar N* state, we show step-by-step derivations of $\widetilde{K}$ and $\widetilde{B}$ for both the N* and HN* states in the following.

As a general approach, we separately derive the expressions for $\widetilde{K}$ and $\widetilde{B}$, corresponding to focusing on either the bending or the dilation of quasi-layers (Fig. 3c). By equalizing Oseen-Frank and Lubensky-de Gennes expressions, $\widetilde{K}$ and $\widetilde{B}$ will be obtained. In the N* state, the polar terms are dropped out in the Oseen-Frank expression. Let us consider a helical director field with its helical axis along the z-axis, i.e., $\mathbf{n} = (\cos qz, \sin qz, 0)$. $q$ is the wave number of the actual pitch length ($q = 2\pi/p$). For the pure dilation of quasi-layers ($q \neq q_0$), only the twist elastic term in the Oseen-Frank form ($f_{\text{O-F}}$) and the compression term in the Lubensky-de Gennes form ($f_{\text{L-G}}$) remain. The energy rises from both terms are:

$$f_{\text{O-F}} = \frac{s^4 K_{22}}{2}(q - q_0)^2 \quad (18)$$

$$f_{\text{L-G}} = \frac{\widetilde{B}}{2}\left(\frac{p}{p_0} - 1\right)^2 \quad (19)$$

Then, we obtain $\widetilde{B} = s^4 K_{22} q_0^2$ for the N* state. The effective quasi-layer compression coefficient $\widetilde{B}$ is related to the equilibrium pitch of the helical director fields. In the high-order-parameter limit of $s = 1$, $\widetilde{B} = K_{22} q_0^2$, consistent with the classic solution[42, 43]. Meanwhile, to derive the expression of $\widetilde{K}$, a pure bending mode of the quasi-layers is considered (also called a jelly-roll arrangement of quasi-layers[42]). In cylindrical coordinates $(r, \phi, z)$, the corresponding director field can be written as:

$$n_r = 0 \tag{20}$$
$$n_\phi = \cos[\theta(r)] \tag{21}$$
$$n_z = \sin[\theta(r)] \tag{22}$$

$\theta$ is the azimuth angle in the plane perpendicular to the $r$-axis. Since the pitch of the system does not change under the pure bending mode, the condition of $\theta(r + p_0) \equiv \theta(r)$ has to be met. Then, only the elastic term in the Lubensky-de Gennes form survives, i.e.,

$$\Delta f_{\text{L-G}} = \frac{\widetilde{K}}{2r^2}. \tag{23}$$

In the Oseen-Frank form, the corresponding free energy term is

$$\Delta f_{\text{O-F}} = \frac{s^4 K_{22}}{2}\left(\frac{d\theta}{dr} - q_0 - \frac{\sin 2\theta}{2r}\right) + \frac{s^4 K_{33}}{2r^2}\cos^4\theta \tag{24}$$

When the bending deformation is just at the onset with an infinite radius of curvature, the approximation $q_0 r \gg 1$ is valid. In the near equilibrium state, the first term in Eq. (24) can be ignored due to $\theta \approx q_0 r + \text{const}$. Since the average over angles of $\cos^4\theta$ is 3/8, so we reach $\widetilde{K} = 3s^4 K_{33}/8$ for the N* state. This scenario was well textbooks[42].

In the HN* state, the polar terms in the extended Oseen-Frank expression come into play. If a pure dilation mode occurs in a homogeneous HN* state ($s$ is constant everywhere), the energy cost

$$f_{\text{O-F}} = \frac{s^4 K_{22}}{2}(q - q_0)^2 + \frac{h}{2}s^2 P_0^2 q^2. \tag{25}$$

Thus, the wave number of the equilibrium pitch for the HN* state becomes

$$q_e = \frac{s^2 K_{22}}{s^2 K_{22} + h P_0^2} q_0 \tag{26}$$

Then, one can easily see that the equilibrium pitch length of the HN* state is elongated by the polarisation gradient interaction ($q_e < q_0$), as illustrated in Fig. 3d. We already know that the effective quasi-layer compression coefficient $\widetilde{B}$ is proportional to the quadratic of the wave number of the equilibrium pitch. Therefore, we obtain

$$\widetilde{B} = \frac{K_{22} q_0^2}{(1 + \kappa)^2} \tag{27}$$

for the HN* state. $\kappa$ is the radio of polarisation gradient interaction and twist elastic modulus, i.e., $\kappa = hP_0^2/s^2 K_{22}$. If the system is near the iso state ($s \to 0$), then the effective layer compression coefficient $\tilde{B}$ becomes zero. For the pure bending mode of the quasi-layers in the HN* state, the polar interaction also contributes to the energy cost, which can be written as

$$\Delta f_{\text{O-F}} = \frac{s^4 K_{33} \cos^4 \theta}{2r^2} + \frac{h}{2} s^2 P_0^2 \left( \frac{q_0 \sin 2\theta}{r} + \frac{\sin^2 \theta \cos^2 \theta}{r^2} \right) \tag{28}$$

The average over angles of $\cos^4 \theta$, $\sin 2\theta$ and $\sin^2 \theta \cos^2 \theta$ are 3/8, 0 and 1/8, respectively. Compared with Eq. (23), the effective elastic coefficient for the quasi-layer of the HN* state is

$$\tilde{K} = \frac{hs^2 P_0^2 + 3s^4 K_{33}}{8} \tag{29}$$

The gradient term, i.e., considering both the Landau and depolarisation effects, enhances the resistance to the splay deformation of the polarisation field.

*Conclusions and Outlooks*

In summary, we start from the typical Oseen-Frank free-energy functionals and arrive at the generalized free-energy functionals for the emerging $N_F$ and HN* LC states. We find that, following the symmetry argument, the odd-rank term of the nematic elasticity arises in polar LCs, corresponding to the polarity-elasticity coupling. In addition, we introduce a necessary coupling between the scalar order parameter and the polarisation order in the free energy functionals. This coupling is critical to conduct theoretical analyses and numerical simulations. Also, in the chiral and polar helices, the polarity directly couples to the effective nematic elasticity and the compressive nature of the quasi-layer systems. The theoretical foundation sets a new challenge in predicting new polar topology and their stability and enables a comparison between theory and experimental measurements of elastic and polar properties.

*Conflicts of interest*

There are no conflicts to declare.

*Acknowledgments*

S.A. acknowledges the supports from the National Key Research and Development Program of China (No. 2022YFA1405000), the Research Fund for International Excellent Young Scientists (RFIS-II; No. 1231101194), the International Science and Technology Cooperation Program of Guangdong province (No.2022A0505050006), the Fundamental Research Funds for the Central University (No. 2022ZYGXZR001), the Recruitment Program of Guangdong (No. 2016ZT06C322) and the 111 Project (No. B18023).

*Figures*

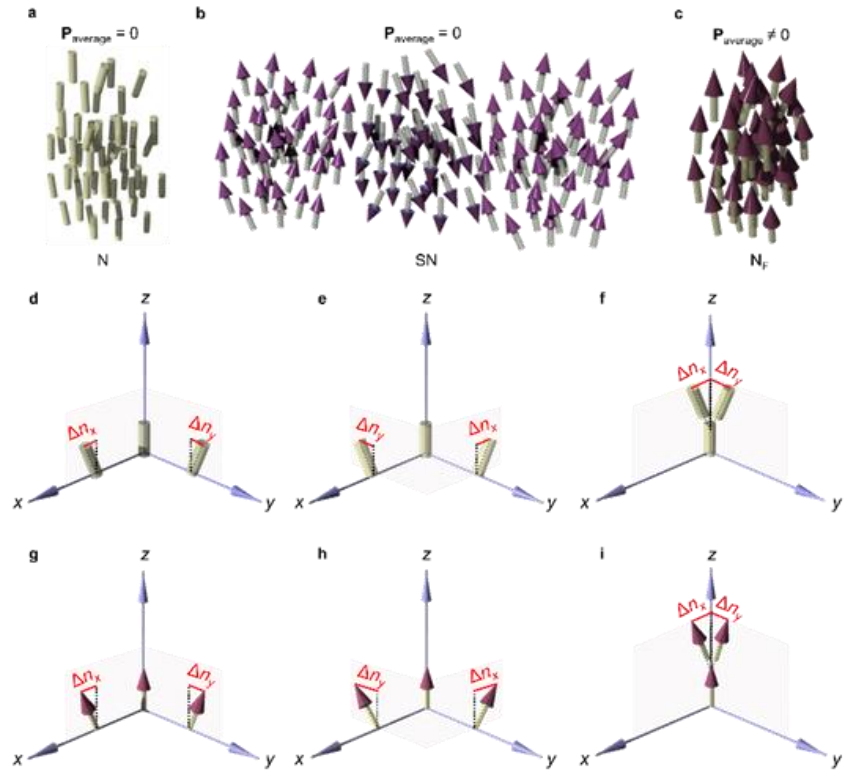

**Fig. 1.** Schematics of the N (a), SN (b) and $N_F$ states (c). The deformation modes for the N (d-f) and NF (g-i) states: splay mode (d,g), twist mode (e,h), and bend mode (f,i).

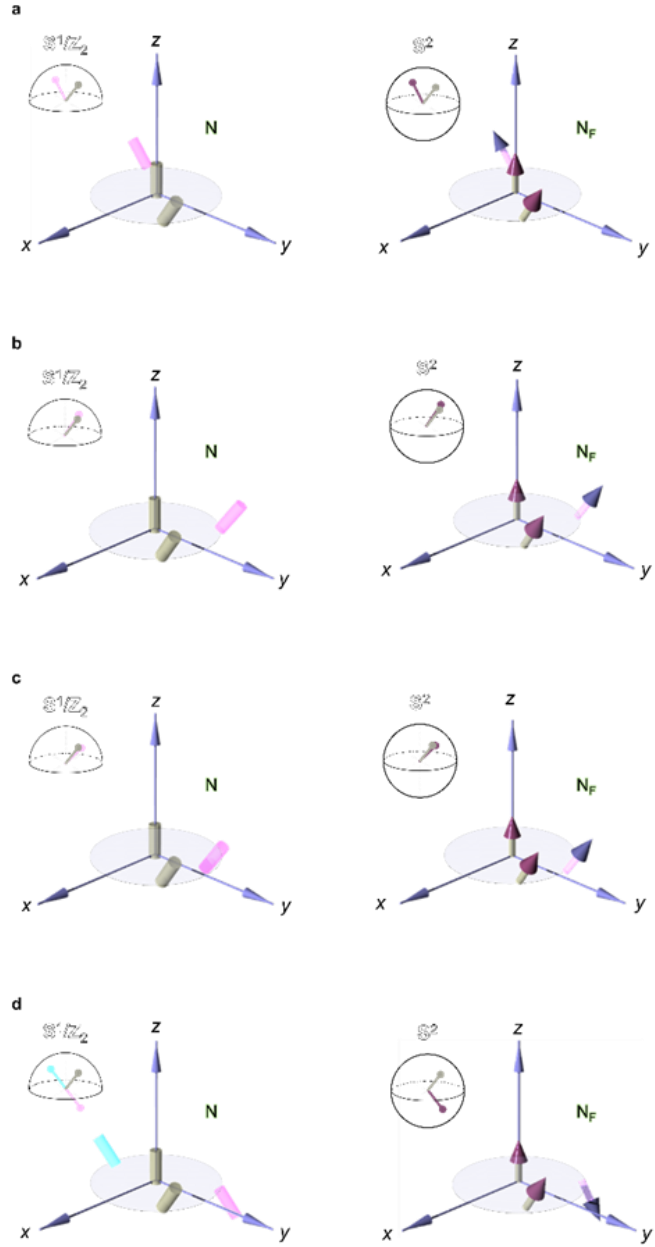

**Fig. 2.** Variation of the director under coordinate transformation in the N (as a reference) and $N_F$ states: $\pi$ rotation about $z$-axis (a), $\pi/2$ rotation about $z$-axis (b), $\pi/4$ rotation about $z$-axis (c), and $\pi$ rotation about $x$-axis (d). The purple rods are the directors after the coordinate rotations. The corresponding coordinates before and after the coordinate rotations in the $S^1/Z_2$ and $S^2$ order parameter spaces are displayed for each coordinate transformation. In (d), the orientational state after the coordinate transformation (pink rod and pink point in the order parameter space) should be remapped to the orientational state (blue rod and blue point in the order parameter space).

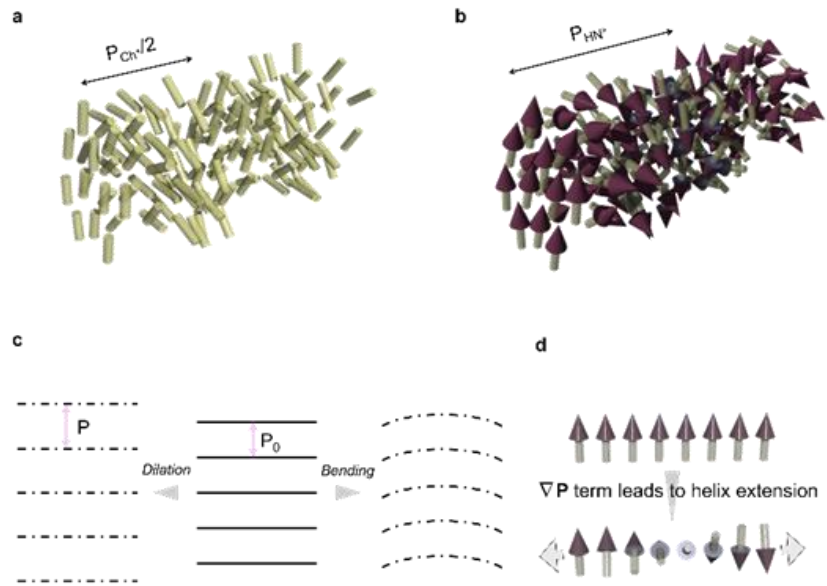

**Fig. 3.** The schematics of the N* (a) and HN* (b) states. (c) Dilation and bending deformation of N* and HN* quasi-layers. (d) The gradient polarisation energy prefers a uniform ferroelectric domain and tends to avoid the helix.